# A superconducting praseodymium nickelate with infinite layer structure

*Motoki Osada,*[\*,†,‡] *Bai Yang Wang,*[‡,§] *Berit H. Goodge,*[∥] *Kyuho Lee,*[‡,§] *Hyeok Yoon,*[‡,⊥] *Keita Sakuma,*[#] *Danfeng Li,*[‡,⊥] *Masashi Miura,*[⊥,#] *Lena F. Kourkoutis,*[∥,∇] *and Harold Y. Hwang*[\*,‡,⊥]

[†]Department of Materials Science and Engineering, Stanford University, Stanford, CA 94305, United States.

[‡]Stanford Institute for Materials and Energy Sciences, SLAC National Accelerator Laboratory, Menlo Park, CA 94025, United States.

[§]Department of Physics, Stanford University, Stanford, CA 94305, United States.

[∥]School of Applied and Engineering Physics, Cornell University, Ithaca, NY 14853, United States.

[⊥]Department of Applied Physics, Stanford University, Stanford, CA 94305, United States.

[#]Graduate School of Science & Technology, Seikei University, Musashino, Tokyo 180-8633, Japan.

[∇]Kavli Institute at Cornell for Nanoscale Science, Cornell University, Ithaca, NY 14853, United States.

ABSTRACT: A variety of nickel oxide compounds have long been studied for their manifestation of various correlated electron phenomena. Recently, superconductivity was observed in nanoscale infinite layer nickelate thin films of $Nd_{0.8}Sr_{0.2}NiO_2$, epitaxially stabilized on $SrTiO_3$ substrates via topotactic reduction from the perovskite precursor phase. Here we present the synthesis and properties of $PrNiO_2$ thin films on $SrTiO_3$. Upon doping in $Pr_{0.8}Sr_{0.2}NiO_2$, we observe superconductivity with a transition temperature of 7-12 K, and robust critical current density at 2 K of 334 kA/cm$^2$. These findings indicate that superconductivity in the infinite layer nickelates is relatively insensitive to the details of the rare earth 4*f* configuration. Furthermore, they motivate the exploration of a broader family of compounds based on two-dimensional $NiO_2$ planes, which will enable systematic investigation of the superconducting and normal state properties and their underlying mechanisms.

The first demonstration of the synthesis of the infinite layer nickelate structure was in $LaNiO_2$ by Crespin *et al.*,[1,2] using hydrogen gas to de-intercalate oxygen from polycrystalline $LaNiO_3$. Subsequently, Hayward *et al.* reported the synthesis of powder $LaNiO_2$ and $NdNiO_2$ using metal hydrides for topotactic reduction, suggesting a paramagnetic response with little (intrinsic) temperature dependence of magnetic susceptibility.[3,4] Further studies showed that nanometer scale single crystalline $LaNiO_2$ thin films could be stabilized in the vicinity of a $SrTiO_3$ substrate, and that they exhibited semiconducting behavior with relatively weak temperature dependence in the resitivity.[5,6] Recently, however, Li *et al.* observed superconductivity in infinite layer neodymium nickelate upon doping with strontium ($Nd_{0.8}Sr_{0.2}NiO_2$), where the investigation of the Nd-compound was motivated to increase the electronic bandwidth with respect to LaNiO$_2$, by virtue of the smaller ionic radius of Nd versus La.[7] This result has renewed long-standing questions regarding the relationship between the electronic structure of infinite layer nickelates and cuprates, particularly with regards to superconductivity, where both cases are derived via doping from a transition metal 3$d^9$ configuration.[8–14]

Superconductivity in cuprates occurs in a wide variety of crystal structures,[15–20] which include octahedral, pyramidal, or square-planar geometries of Cu-O. In all cases, the common and essential feature is the two-dimensional (2D) CuO$_2$ plane, around which a wide range of rare-earth (*RE*) substitutions can be accommodated. In this context, it is reasonable to consider the possible existence of other compounds in a broader family of superconducting nickelates, with 2D NiO$_2$ planes as the common structural element. As a first step, praseodymium (*RE* = Pr in *RE*NiO$_2$) is a reasonable starting point, since it is already known that for the quasi-2D reduced trilayer nickelate Pr$_4$Ni$_3$O$_8$, the ground state is metallic, in contrast to the density-wave insulator found for La$_4$Ni$_3$O$_8$.[21,22] Thus it may be hoped that upon doping, a metallic and ultimately superconducting state can be stabilized in PrNiO$_2$. Pr is also interesting from the viewpoint of testing possible underlying aspects of the electronic structure that may be pertinent for superconductivity in Nd$_{0.8}$Sr$_{0.2}$NiO$_2$. For example, it has recently been suggested that direct hybridization with the Nd 4$f$ states may be important for the electronic structure of NdNiO$_2$ near the chemical potential.[23] This would require a significant degree of fine-tuning of the electron interaction strength acting on these 4$f$ states, which would be unlikely to occur also for other rare-earth compounds.

It should be noted that there are no prior reports of PrNiO$_2$ in either bulk or thin film form. In considering its synthesis, since the crystallinity of the precursor perovskite phase has been shown to strongly affect the reduction process,[3,24] growth of highly crystalline PrNiO$_3$ is a necessary first step. The structure of an ideal perovskite oxide $ABO_3$ can be described using the tolerance factor, where $A$ and $B$ are the cationic sites. The tolerance factor reaches 1 when the three ionic spheres of $A^{3+}$, $B^{3+}$ and $O^{2-}$ fit perfectly. In the case of $RE$NiO$_3$, the tolerance factor decreases as the atomic number of $RE$ increases, leading to an increasingly distorted NiO$_6$ octahedral network that governs the overall electronic and magnetic phase diagram.[25] Note that unlike the infinite layer nickelates, in distorted perovskite nickelates reducing the $RE$ ionic radius decreases the electronic bandwidth, by decreasing the Ni $3d$ – O $2p$ – Ni $3d$ overlap integral. Since the tolerance factor for PrNiO$_3$ lies between that for NdNiO$_3$ and LaNiO$_3$, conditions for thin film synthesis should also be intermediate between these two points of reference.[26] Here we report the synthesis of infinite-layer praseodymium nickelate thin films on SrTiO$_3$, and by doping with strontium (Pr$_{0.8}$Sr$_{0.2}$NiO$_2$), we observe a superconducting transition at 7 – 12 K.

Precursor perovskite thin films were grown using pulsed laser deposition (PLD) from polycrystalline targets ablated with a KrF excimer laser (wavelength 248 nm). Prior to their growth, SrTiO$_3$ (001) substrates were pre-annealed at 910 °C under an oxygen partial pressure of $5\times10^{-6}$ Torr, in order to obtain an atomically flat surface topography with unit cell step-and-terrace structure. Here, SrTiO$_3$ (001) was used because the estimated in-plane lattice mismatch between PrNiO$_2$ and SrTiO$_3$ is minimized among the commonly available perovskite substrates. During the deposition, the oxygen partial pressure was 200 mTorr for PrNiO$_3$ and 250 mTorr for Pr$_{0.8}$Sr$_{0.2}$NiO$_3$; the substrate temperature was 570 °C for both. The laser was operated at a fluence of 1.39 J/cm$^2$ for PrNiO$_3$ and 2.19 J/cm$^2$ for Pr$_{0.8}$Sr$_{0.2}$NiO$_3$, with a repetition rate of 4 Hz. We used a uniform 2.64 mm$^2$ laser spot for ablation, formed by imaging an aperture, which was previously found to be important for improving sample-to-sample reproducibility for Nd$_{0.8}$Sr$_{0.2}$NiO$_2$ films.[24]

Unlike these prior studies, where an upper SrTiO$_3$ capping layer was essential for stabilizing the infinite layer structure in (Nd,Sr)NiO$_2$, we found that PrNiO$_3$ and Pr$_{0.8}$Sr$_{0.2}$NiO$_3$ could be stably reduced up to a thickness of around 12 nm. Thus the nickelate films were primarily synthesized without a SrTiO$_3$ capping layer; for comparison, we also prepared a sample (Pr$_{0.8}$Sr$_{0.2}$NiO$_2$) with a 2 nm SrTiO$_3$ cap. For SrTiO$_3$ deposition, the oxygen partial pressure, the substrate temperature, and the laser fluence were 200 mTorr, 570 °C, and 0.90 J/cm$^2$, respectively, with a repetition rate of 4 Hz. As-grown films were placed in a Pyrex glass tube with calcium hydride (CaH$_2$) powder (~ 0.1 g), with the films loosely wrapped with aluminum foil to avoid direct physical contact with the powder. The glass tube was sealed using a hydrogen torch while under vacuum below 0.1 mTorr using a rotary pump. In a tube furnace, the sealed glass tube was annealed at 210 – 240 °C for 45 – 60 min for uncapped films and at 260 °C for 60 min for films with a 2 nm SrTiO$_3$ cap, where the ramping and cooling rates were 10 °C/min.

Samples before and after reduction were characterized by X-ray diffraction (XRD) $\theta$–2$\theta$ symmetric scans and reciprocal space mapping (RSM) using monochromated Cu $K\alpha_1$ radiation. Cross-sectional scanning transmission electron microscopy (STEM) specimens were prepared using a standard focused ion beam (FIB) lift-out process on a Thermo Scientific Helios G4 X FIB. Samples were imaged on an aberration-corrected FEI Titan Themis operated at 300 keV with a 30 mrad probe-forming convergence angle. Electron energy-loss spectroscopy was performed on the same Titan Themis system equipped with a 965 GIF Quantum ER and a Gatan K2 Summit direct electron detector operated in electron counting mode. The electrical resistivity of the 2.5 mm x 5 mm films was measured using a four-point geometry using Al wire bonded contacts. The current-voltage (*I-V*) measurements were performed on a narrow channel of 0.1 mm width. Here standard photolithography was used, and the film was wet-etched with 5.5% aqueous solution of nitric acid for 7 seconds and rinsed with deionized water. The remaining photoresist was removed by acetone and isopropyl alcohol. We confirmed that the superconducting transition and resistivity were unchanged before and after processing. Mutual inductance was measured using two coils in a collinear configuration. The twin 80-turn coils had outer diameter of ~ 1.5 mm and inner diameter of ~ 0.5 mm, yielding a self-inductance of ~ 6 µH. The drive coil was driven with an alternating current of root-mean-square amplitude of 100 µA and frequency of 15 kHz. The in-phase and out-of-phase components of the voltage across the pick-up coil were measured by lock-in amplification. The measured voltage was in a regime of linear response with respect to the amplitude of the drive current.

Figure 1a,b show XRD $\theta$–$2\theta$ symmetric scans of as-grown $PrNiO_3$ and $Pr_{0.8}Sr_{0.2}NiO_3$, and after topotactic reduction. The as-grown films show clear pseudocubic perovskite 00$l$ peaks with robust 001 peak intensity, and no other features, indicating single-phase perovskite epitaxial thin films. After reduction, only peaks corresponding to infinite layer nickelate 00$l$ are observed. The out-of-plane lattice constant is 3.31 Å (3.39 Å) for infinite-layer $PrNiO_2$ ($Pr_{0.8}Sr_{0.2}NiO_2$). To probe the in-plane lattice constant and strain state, RSM was performed around the $SrTiO_3$ $\bar{1}03$ peak, as shown in Figure 1c,d for $Pr_{0.8}Sr_{0.2}NiO_3$ and $Pr_{0.8}Sr_{0.2}NiO_2$. In all samples, both doped and undoped as well as before and after reduction, the films are epitaxially strained to the $SrTiO_3$ substrate with an in-plane lattice constant of 3.91 Å. Considering the larger ionic radius of Pr, the natural $a$-axis parameter of infinite-layer $PrNiO_2$ is expected to be slightly larger than that reported for bulk $NdNiO_2$ (3.92 Å),[3] resulting in compressive in-plane strain for all of the films studied.

Figure 2a,b display cross-sectional high-angle annular dark field (HAADF) and annular bright field (ABF) STEM images of a $Pr_{0.8}Sr_{0.2}NiO_2$ film on $SrTiO_3$ substrate, respectively, showing the formation of the infinite-layer structure with $NiO_2$ square-planar geometry. In particular, the removal of O from the Pr/Sr planes oriented parallel to the substrate is clear in the ABF image, confirming successful reduction of the film to the infinite-layer phase. Figure 2c shows an annular dark field (ADF) STEM image, with similarly coherent structure observed especially in the vicinity of the interface. Some crystalline defects can be seen that are qualitatively similar to those observed in $Nd_{0.8}Sr_{0.2}NiO_2$.[24] The elemental maps in Figure 2d,e,f show corresponding elemental contrast from the integrated Ti-$L_{2,3}$, Ni-$L_{2,3}$, and Pr-$M_{4,5}$ edges, respectively, confirming an abrupt interface with the substrate and generally uniform distribution of Pr and Ni throughout the film (Figure 2g).

The temperature-dependent resistivity $\rho(T)$ of perovskite $PrNiO_3$ and $Pr_{0.8}Sr_{0.2}NiO_3$ are plotted in Figure 3a. $PrNiO_3$ displays a clear hysteretic metal-insulator transition at 130 K ($T_{MI}$, warming transition), indicative of a first-order transition with the resistivity changing over one order of magnitude. This result is in good agreement with previous reports for thin film samples.[26] Compared with $NdNiO_3$ ($T_{MI}$ = 180 – 210 K),[7,26] $T_{MI}$ for $PrNiO_3$ shifts to lower temperature due to a reduction of the charge transfer gap by the increased tolerance factor in the $PrNiO_3$ lattice, which increases the electronic bandwidth.[25] For $Pr_{0.8}Sr_{0.2}NiO_3$, the resistivity decreases to 0.1 mΩ·cm, showing metallic temperature dependence consistent with the doping suppression of $T_{MI}$ previously reported.[27,28] The behavior of $\rho(T)$ after reduction to the infinite layer structure is shown in Figure 3b. Both in magnitude and detailed temperature dependence, $PrNiO_2$ is remarkably similar to $NdNiO_2$, and has a minimum resistivity of ~ 1.3 mΩ·cm with a low temperature upturn below ~ 70 K. By contrast, $Pr_{0.8}Sr_{0.2}NiO_2$ is overall more conductive, and exhibits a superconducting transition with a zero-resistance temperature $T_{c,\,zero}$ of 7 K and an onset temperature $T_{c,\,onset}$ of 12 K (defined as the temperature at which the resistivity is 90% of the value at 20 K). Here we show five different $Pr_{0.8}Sr_{0.2}NiO_2$ samples with thicknesses ranging from 5.3 – 12 nm, in one case with a 2 nm SrTiO$_3$ cap. As recently observed for (Nd,Sr)NiO$_2$, $T_c$ is quite reproducible, while the magnitude of the normal state resistivity has considerable variation.[29,30] We attribute the former to the use of precise imaging conditions and re-polishing the target for each growth, and the latter to sample-to-sample variations in the extended defect structure of the films. Across the thickness range studied, no systematic variation in $T_c$ or normal state resistivity was observed (Figure 3c,d). Overall these values for $T_c$ are comparable to that for Nd$_{0.8}$Sr$_{0.2}$NiO$_2$, and somewhat lower than the highest observed.[7,24]

The $I$-$V$ characteristics in the superconducting state for a patterned device (Figure 4a) are shown in Figure 4b at several temperatures. At low temperatures, a dissipationless current is observed below the critical current density $J_c$, followed by ohmic (linear) behavior. The initial temperature dependence of $J_c$ is shown in the inset of Figure 4b, where $J_c(T)$ is estimated as the maximum in d$V$/d$I$. It is noteworthy that the 2 K value for $J_c$ is twice as large as previously observed for Nd$_{0.8}$Sr$_{0.2}$NiO$_2$.[7] To investigate the diamagnetic response of the superconducting film, we performed a two-coil mutual-inductance measurement with the $c$-axis aligned along the longitudinal axis of the coil, as shown in Figure 4c. Below $T_{c,\text{zero}}$, the real (in-phase) component of $V_p$, the voltage across the pick-up coil, drops and the imaginary (out-of-phase) part of $V_p$ shows a peak, corresponding to diamagnetic screening when the film becomes (globally) superconducting. Similar to Nd$_{0.8}$Sr$_{0.2}$NiO$_2$, these results indicate a London penetration depth on the micron scale.[7]

In summary we have observed superconductivity in doped Pr$_{0.8}$Sr$_{0.2}$NiO$_2$, providing a first step in expanding the family of the infinite-layer nickelate superconductors. In terms of the questions that motivated this exploration, the observation of a nearly comparable $T_c$ in Pr$_{0.8}$Sr$_{0.2}$NiO$_2$ to Nd$_{0.8}$Sr$_{0.2}$NiO$_2$ suggests that fine-tuning of a direct 4$f$ hybridization to states near the Fermi level is not an essential requirement for superconductivity, as it is unlikely for this to occur for both $Nd_{0.8}Sr_{0.2}NiO_2$ and $Pr_{0.8}Sr_{0.2}NiO_2$. Specific $Nd^{3+}$ fluctuations to its secondary oxidation state $Nd^{2+}$, possibly analogous to heavy fermion systems,[31–34] also seem unnecessary, since for Pr the secondary oxidation state is 4+. Hybridization between rare-earth 5$d$ states and Ni 3$d$ states appears equally relevant for both cases.[35] The role, if any, of 4$f$ magnetism with respect to superconductivity is unknown at this time. While $La^{3+}$ has an electronic configuration of $[Xe]4f^0$ with no magnetic moment, $Pr^{3+}$ and $Nd^{3+}$ have $[Xe]4f^2$ and $[Xe]4f^3$ configurations with local magnetic moments, respectively. Finally, while we benefitted from insights of a recent study optimizing the growth parameters of $Nd_{0.8}Sr_{0.2}NiO_2$,[24] we have not yet performed a comparably extensive study for $Pr_{0.8}Sr_{0.2}NiO_2$. Nevertheless our initial impression is that this is a more forgiving materials system than $Nd_{0.8}Sr_{0.2}NiO_2$. For example, even without a supporting $SrTiO_3$ capping layer, we could stabilize the infinite layer structure (up to ~ 12 nm) with a thickness (estimated from Scherrer fits to XRD)[24] as high as ~ 96% of the total thickness deduced from x-ray reflectivity measurements. This likely reflects the larger tolerance factor and closer lattice match to the substrate in the perovskite precursor phase, suggesting that $(Pr,Sr)NiO_2$ compounds may provide an avenue to continue to refine the crystallinity accessible in infinite-layer nickelate superconductors.

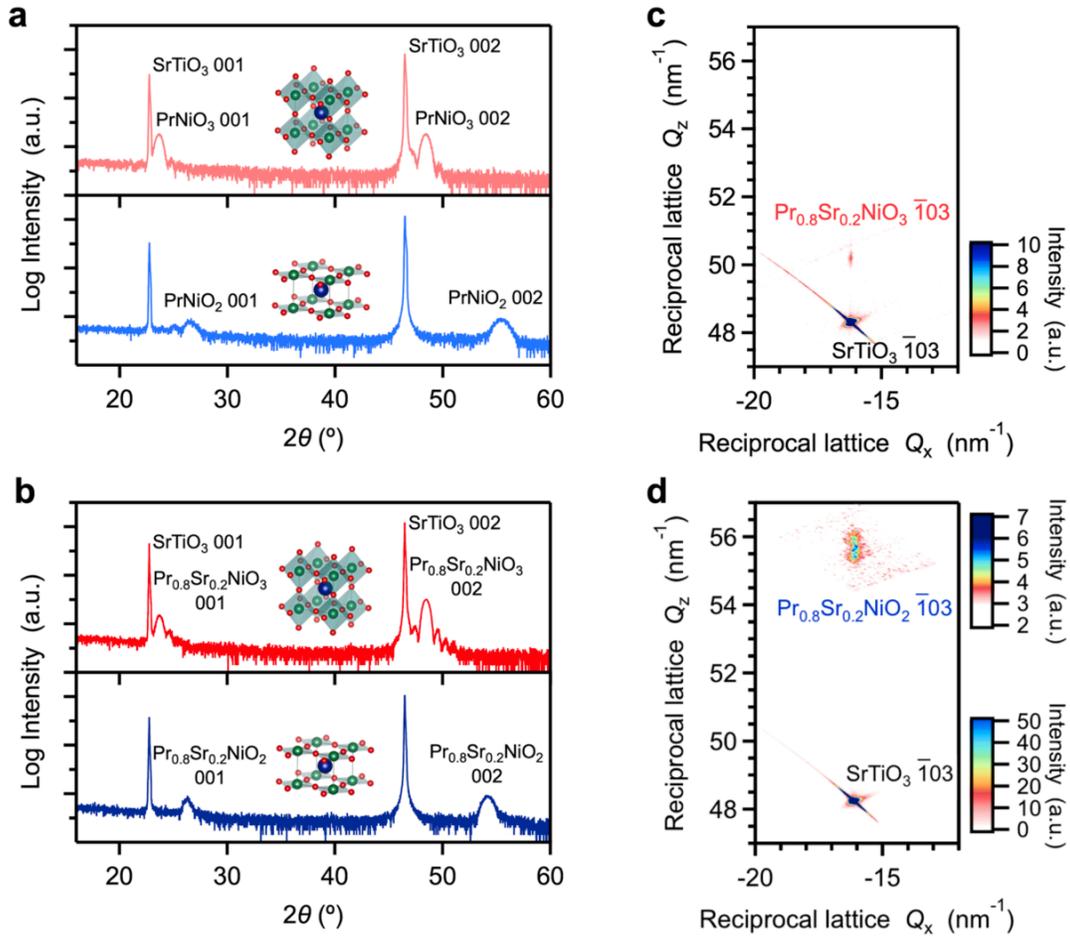

**Figure 1.** X-ray diffraction $\theta$–$2\theta$ symmetric scans of (a) undoped thin films (PrNiO$_3$ and PrNiO$_2$), and (b) doped thin films (Pr$_{0.8}$Sr$_{0.2}$NiO$_3$ and Pr$_{0.8}$Sr$_{0.2}$NiO$_2$) grown on SrTiO$_3$ (001) substrates. The atomic structure of the perovskite and infinite-layer phases are shown as inserts. Reciprocal space maps of (c) Pr$_{0.8}$Sr$_{0.2}$NiO$_3$, and (d) Pr$_{0.8}$Sr$_{0.2}$NiO$_2$ around the $\bar{1}03$ peaks.

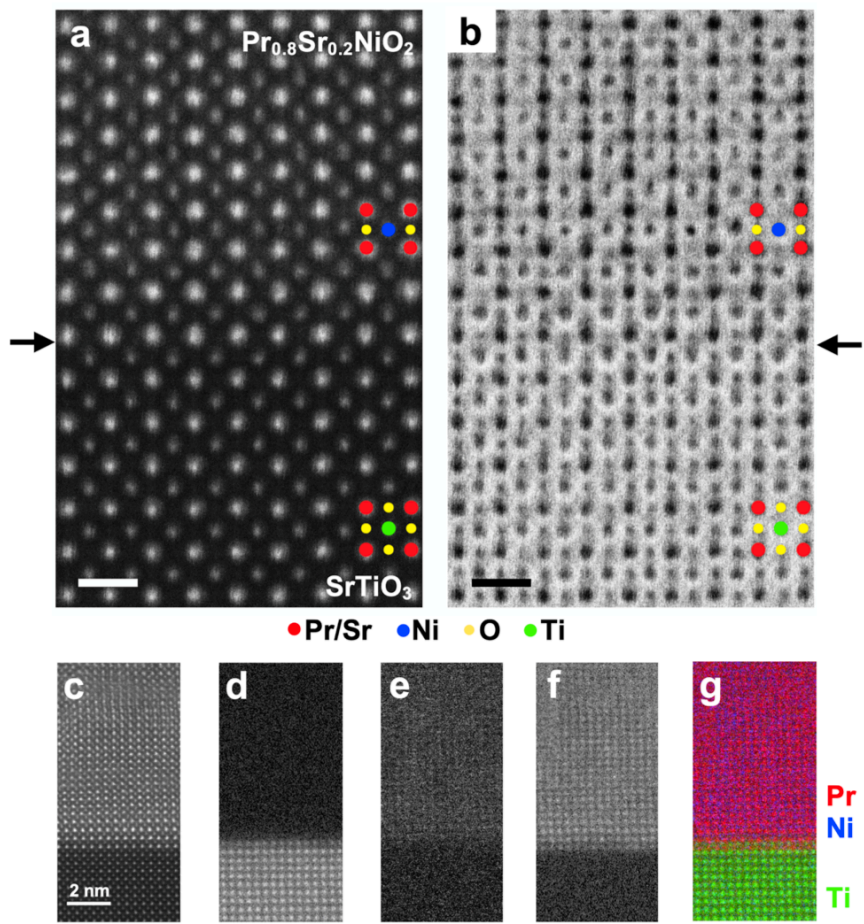

**Figure 2.** Cross-sectional scanning transmission electron microscopy (STEM) images of $Pr_{0.8}Sr_{0.2}NiO_2$ on a $SrTiO_3$ substrate. (a) High-angle annular dark field (HAADF) STEM image of the interface region. (b) Annular bright field (ABF) STEM image of the same region showing successful O-reduction to the infinite-layer phase. Red dots show Pr/Sr; blue, yellow, and green dots represent Ni, O, and Ti, respectively. Arrows indicate the interface. Scale bars denote 5 Å. (c) Annular dark field (ADF) STEM image of $Pr_{0.8}Sr_{0.2}NiO_2$ on $SrTiO_3$ and simultaneously-acquired elemental maps of the (d) Ti-$L_{2,3}$ edge, (e) Ni-$L_{2,3}$ edge, and (f) Pr-$M_{4,5}$ edge of the same region in (c). (g) False-colored elemental map with Pr, Ni, and Ti shown in red, blue, and green, respectively.

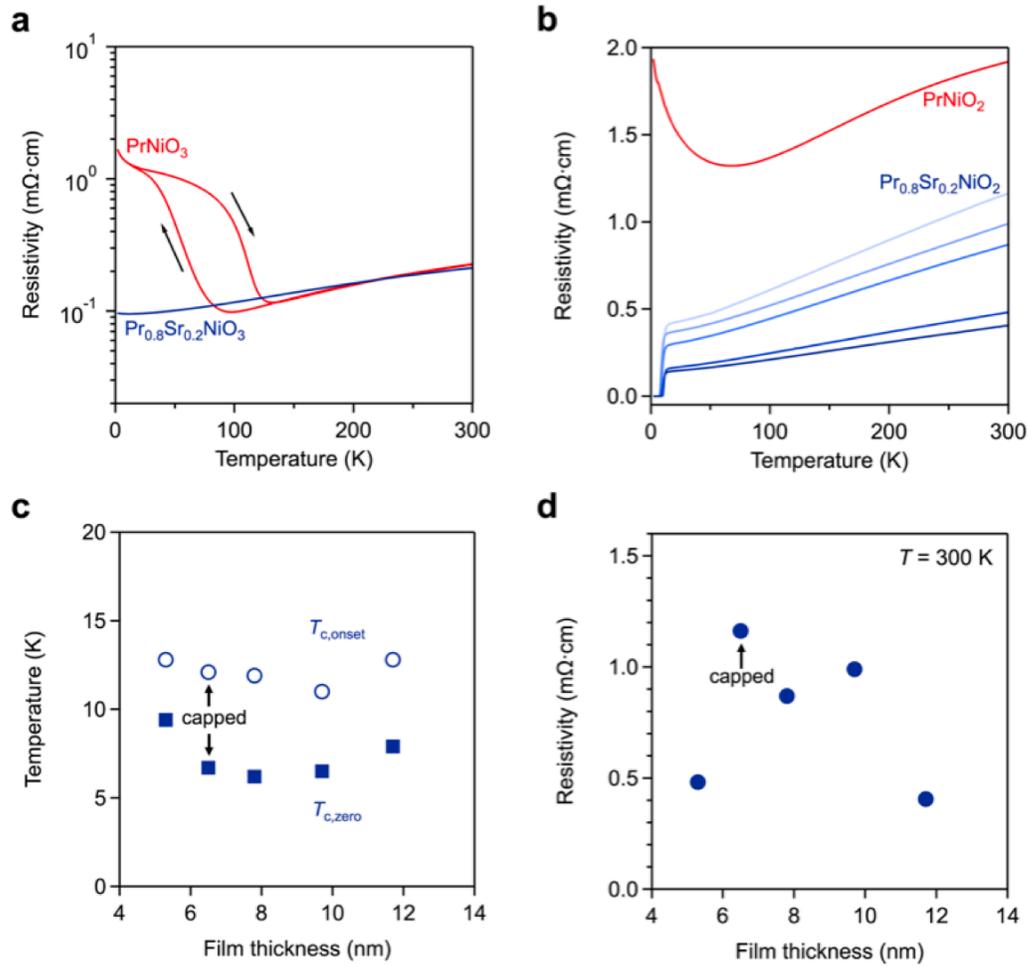

**Figure 3.** Temperature dependent resistivity for undoped and doped (a) as-grown perovskite films (PrNiO$_3$ and Pr$_{0.8}$Sr$_{0.2}$NiO$_3$), and (b) reduced infinite layer films (PrNiO$_2$ and Pr$_{0.8}$Sr$_{0.2}$NiO$_2$). Arrows indicate cooling and warming cycles. Data taken on five Pr$_{0.8}$Sr$_{0.2}$NiO$_2$ films 5.3 – 12 nm thick grown under the same conditions are shown (blue). The top curve is a film capped with 2 nm of SrTiO$_3$, and the other four curves correspond to uncapped films. Thickness dependence of (c) the superconducting transition temperature, and (d) room temperature resistivity.

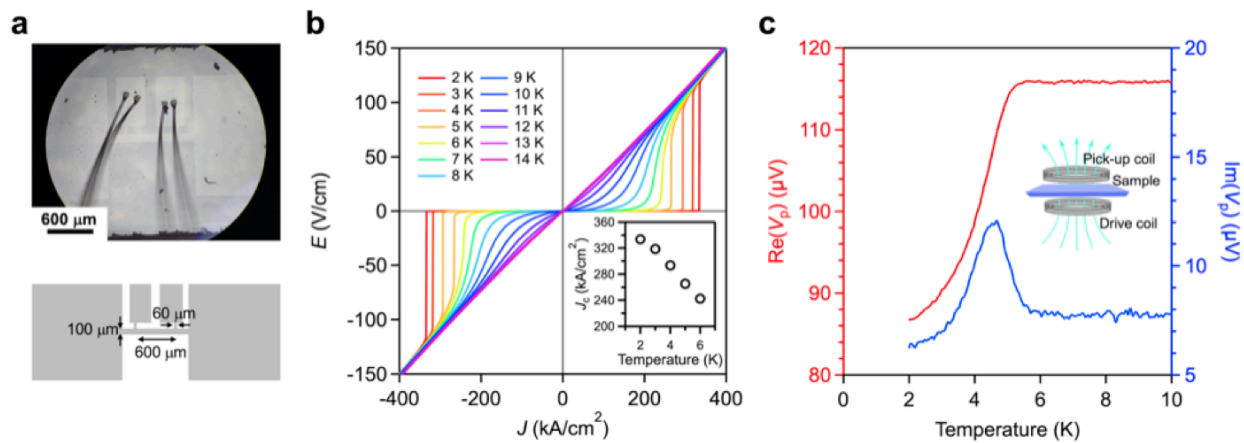

**Figure 4.** (a) Optical microscope image (top) and schematic (bottom) of the lithographically-patterned device used for *I-V* measurements. (b) Electric field (*E*) versus current density (*J*) characteristics for $Pr_{0.8}Sr_{0.2}NiO_2$, taken from 2 K to 14 K. The inset shows the critical current density $J_c$ as a function of temperature. (c) Two-coil mutual inductance response of a $Pr_{0.8}Sr_{0.2}NiO_2$ film as a function of temperature. The real (in-phase) part, $Re(V_p)$, and imaginary (out-of-phase) part, $Im(V_p)$, of the signal are shown on the left and right axes, respectively. The inset shows a schematic diagram of the experiment.

AUTHOR INFORMATION

**Corresponding Authors**

Motoki Osada; mosada@stanford.edu

Harold Y. Hwang; hyhwang@stanford.edu

**Notes**

The authors declare no competing financial interest.

ACKNOWLEDGMENTS

This work was supported by the US Department of Energy, Office of Basic Energy Sciences, Division of Materials Sciences and Engineering, under contract number DE-AC02-76SF00515. M.O. acknowledges partial financial support from the Takenaka Scholarship Foundation. D.L. acknowledges early support by the Gordon and Betty Moore Foundation's Emergent Phenomena in Quantum Systems Initiative through grant number GBMF4415, which also supported synthesis equipment. M.M. acknowledges support by JSPS KAKENHI (18KK0414), the Heiwa Nakajima Foundation, and the Kato Foundation for Promotion of Science (KJ-2744). B.H.G. and L.F.K. acknowledge support by the Department of Defense Air Force Office of Scientific Research (No. FA 9550-16-1-0305). This work made use of the Cornell Center for Materials Research (CCMR) Shared Facilities, which are supported through the NSF MRSEC Program (No. DMR-1719875). The FEI Titan Themis 300 was acquired through No. NSF-MRI-1429155, with additional support from Cornell University, the Weill Institute, and the Kavli Institute at Cornell. The Thermo Fisher Helios G4 X FIB was acquired with support from the National Science Foundation Platform for Accelerated Realization, Analysis, and Discovery of Interface Materials (PARADIM) under Cooperative Agreement No. DMR-1539918.

Abstract Graphics

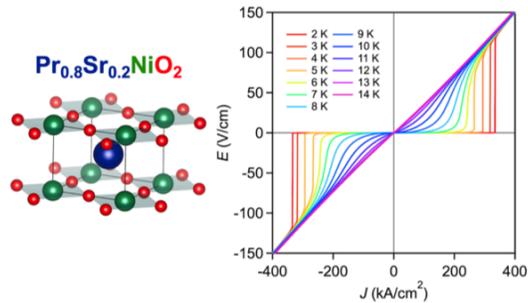